# Observation of Fano resonance using a coupled resonator metamaterial composed of the meta-atoms arranged by double periodicity

Tsubasa Nishida,[1] Yosuke Nakata,[2] Fumiaki Miyamaru,[1,2] Toshihiro Nakanishi,[3] and Mitsuo W. Takeda[1]

*[1] Department of Physics, Faculty of Science, Shinshu University, 3-1-1 Asahi, Matsumoto, Nagano 390-8621, Japan*

*[2] Center for Energy and Environmental Science, Shinshu University, 4-17-1 Wakasato, Nagano 380-8553, Japan*

*[3] Department of Electronic Science and Engineering, Kyoto University, Kyoto 615-8510, Japan*

We studied the transmission characteristics of a planar metamaterial consisting of an array of the electric split-ring resonators (eSRRs) with double periodicity. Because of the coupling between the different resonant modes induced by the different lattice periods in the metamaterials with double periodicity, the appearance of the Fano resonance can be expected in the same manner of the coupled classical oscillators. We fabricated complementary eSRRs and verified that the Fano-like spectral shape appeared in the transmission spectra of the eSRRs with double periodicity in the terahertz region.

In the field of atomic physics, Fano resonance induced by the interference between two excitation passes from two different states shows an asymmetric line in the transmission and reflection spectra.[1] The classical analogue of the Fano resonance using coupled two-harmonic oscillators such as the coupled two resonator circuits through mutual inductance has been reported.[2,3] The interaction between each resonator induces the modulation of both the phase and the amplitude of resonance and results in the sharp resonant line in the transmission spectrum. Metamaterials, composed of an array of artificial sub-wavelength structures called meta-atoms, have been used as for realizing analogues of the Fano resonance.[4-6] The sharp spectral responses of Fano-like metamaterials have been applied to the sensing application[7-9] since the narrow-band resonant line is useful to detect the frequency shift due to the refractive index of sample. Studies of Fano-like metamaterials have been reported in broad-frequency regions such as the infrared,[10,11] terahertz,[12,13] and microwave[14,15] regions because the resonant frequency of metamaterials can be freely tuned by scaling the geometrical parameters of the meta-atoms.

Meta-atoms for Fano resonance are designed to have resonant modes determined by their geometrical parameters. Two different meta-atoms with different geometrical or material parameters are coupled to realize the Fano resonance analogue. In this case, the two resonators have resonant modes with low and high quality factors (Q), and the Fano-like spectrum is observed by the coupling effect between these two different modes.[16]

Although Fano resonance is observed by the coupling of the different resonant modes, Fano-like metamaterials do not necessarily require two different meta-atoms. When the multi-resonant modes that induce Fano resonance in one meta-atom are excited by engineering the parameters of the meta-atom, we can observe the Fano resonance.[17,18] By manipulating the coupling constant between the incident wave and each resonator, the multi-resonant modes can be also excited. Phase-coupled plasmon-induced transparency (PC-PIT) is induced by the phase difference between the incident wave and each resonator.[19] PC-PIT is observed by stacking resonators along the propagation direction of the incident wave[20] or by adjusting the incident angle.[21] When the incident wave has a field gradient, a Fano-like spectrum can be also observed. This transparency is called the field-gradient-induced transparency (FGIT).[22, 23]

Electromagnetic characteristics of many metamaterials are the result of an ensemble of meta-atoms arranged in a sub-wavelength lattice period, and each meta-atom couples to another one through the near-field.[24] The effective resistances or inductances of the meta-atoms include the effect of near-field coupling with respect to the lattice period. Thus, the periodicity of meta-atoms is regarded as a parameter to control the resonant frequency and the line width. In other words, meta-atoms with the different lattice periods induce different resonant modes, even if we use only one type of meta-atom. However, Fano-like metamaterials utilizing the periodicity of meta-atoms have not been reported to our knowledge.

In this work, we demonstrate the Fano-like spectral response of an array of one type of meta-atom with double periodicity. The double periodicity means that one sub-lattice cannot overlap another sub-lattice by a translational operation. Because meta-atoms arranged by each periodicity can be considered different oscillators, Fano resonance is observed by the coupling of these resonant modes. Thus, our proposed metamaterial structure has the following two key features, in contrast to the Fano-like metamaterials of earlier studies. First, only one mode is excited on meta-atoms arranged by single periodicity. Second, we enter a plane wave at normal incidence, and our metamaterial is composed of one layer. Thus, we do not utilize the field gradient and phase difference of the incident wave to induce Fano resonance.

We fabricated the metamaterials with double periodicity using a femtosecond laser ablation process with a 430-nm-thick copper thin film deposited on a 100-µm-thick Zeonor® substrate, as shown in Fig. 1(a). We chose electric split-ring resonators (eSRRs) in order to ignore the excited magnetic dipole for simple analysis. The resonance of this meta-atom originates from counter-circulating currents in the both side of the triangle ring structures. These currents eliminate magnetoelectric response in each unit cell resulting in a pure electric response. To make the fabrication process easier, we used complementary eSRRs (c-eSRRs). The c-eSRRs with double periodicity (cD-eSRRs) were composed of c-eSRRs with sparse (cS-eSRRs) and tight (cT-eSRRs) periodicities, as shown in Figs. 1(b) and (c), respectively. The geometrical parameters of the meta-atoms are shown in Fig. 1(d). The tight and sparse lattice periods were set to 70 µm and 140 µm, respectively. The complementary screen was designed by replacing the metal lines of the positive-type eSRR to the metal slots. Based on Babinet's principle, the field transmission coefficients of the original [$t(\omega)$] and complementary [$t_c(\omega)$] planar metamaterials should satisfy $t(\omega) + t_c(\omega) = 1$. Although our metamaterials were fabricated on a substrate, we can approximately apply Babinet's principle.[25] Then, for inducing the resonant modes, we entered the plane wave that was linearly polarized along the *x* direction at normal incidence, as shown in Fig. 1(d). This polarization was perpendicular to that of the positive-type eSRRs.

We measured the transmitted THz waves through the metamaterials with single and double periodicities by using terahertz time-domain spectroscopy (THz-TDS)[26]. Figure 2 shows the measured THz waves for cT-eSRRs (blue), cS-eSRRs (red), and cD-eSRRs (black). The oscillation time of cS-eSRRs is longer than that of cT-eSRRs. The wave shape of cD-eSRRs until about 18 ps is similar to the cT-eSRRs. The cycle in the interval from 20 ps to 30 ps is very close to the cS-eSRRs.

Figure 3(a) shows the measured power transmission spectra of our samples. The transmission peaks of the cT-eSRRs (blue dotted-dashed line) and the cS-eSRRs (red dashed line) are observed at 0.94 THz and 0.93 THz, respectively. Because the bandwidth of the cS-eSRRs is narrower than that of the cT-eSRRs, the modes of the cT-eSRRs and the cS-eSRRs work as low and high Q resonators, respectively. The estimated Q factors of the cT-eSRRs and the cS-eSRRs are about 9.5 and 23.5, respectively. Because the number of c-eSRRs in the cS-eSRRs is smaller than that in cT-eSRRs, the transmittance of the cS-eSRRs at peak frequency is lower than that of the cT-eSRRs. While the transmission spectra of the cT-eSRRs and the cS-eSRRs show single peaks, two peaks and one dip are observed in the transmission spectrum of the cD-eSRRs (black solid line). The frequency of the transmission dip (reflection window) is close to the resonant frequency of the cS-eSRRs. This spectral response agrees with that of the Fano-like metamaterials.[10,13] Note that we can ignore the diffraction effect because the diffraction frequency is estimated from the lattice period of cS-eSRRs to be approximately 1.40 THz.

In order to verify the experimental results, we also calculated the transmission properties of our samples by the finite element method (COMSOL MULTIPHYSICS®). The unit cell of the cD-eSRRs is shown as the region surrounded by a white dashed line in Fig. 1(a). The complex refractive index of the Zeonor® substrate[27] is $\tilde{N} = 1.53 + 3.00 \times 10^{-3}$i, and the conductivity of copper[28] is $\sigma = 6.00 \times 10^7$ S/m. As shown in Fig. 3(b), the simulated results agree well with the experimental results. The discrepancies can be attributed to the non-uniformity among the resonators due to fabrication tolerances.

Both the experimental and calculated results show Fano-like characteristics in the transmission spectra of the cD-eSRRs. The cT-eSRRs and cS-eSRRs that constitute the cD-eSRRs excite the different single-resonant modes (see Fig. 3). Because they are arranged in a unit cell that is smaller than the resonant wavelength, the two-coupled oscillator model, in which the two LC resonator circuits are coupled by the mutual capacitance, is expected to explain this Fano-like spectral response.

We calculated the spatial distribution of the electric field, $|E|$, of the cD-eSRRs to indicate the validity of the two-oscillator model. Figure 4 shows the calculated distribution of $|E|$ at the two transmission peaks and one dip frequencies [see the black arrows in Fig. 3(b)]. There is no difference for the four resonators at the four corners of the color map. Thus, we can consider this resonant mode as one oscillator. We call this oscillator "TO" because these resonators are arranged by tight periodicity. However, the electric fields of the one resonator at the center of the color map show values different from those of the other four resonators at each frequency. Thus, we can treat this resonator as a different oscillator from the original. We call this oscillator "SO" because these resonators are arranged by sparse periodicity. As shown in Figs. 4(a) and 4(c), all resonators are excited by the incident wave. On the other hand, at the dip frequency, as shown in Fig. 4(b), the electric field is strongly localized in the central resonator. It has been reported that the resonance of a low-Q resonator is hardly excited at the transmission window in studies of coupled metamaterials for the Fano resonance.[12,15] Because we can consider that the reflection window of cD-eSRRs corresponds to the transmission window of coupled metamaterials in earlier studies, the resonant modes of TO and SO correspond to the low and high Q modes, respectively. Because this agrees with the experimental results, that the Q factor of cS-eSRRs, corresponding to SO, is higher than that of cT-eSRRs, corresponding to TO, the localization of the electric field in Fig. 4(b) can be explained by the coupling effect between the low and high Q resonators.

For detailed analysis, we calculated the phase difference of the electric field on each resonator from the phase of the incident wave on same point as shown in Fig. 5. We calculated the components of the electric field parallel to the incident electric field, and the observation points are set to the center of each resonator. When the meta-atoms are aligned with single periodicity, the phase differences for the cT-eSRRs (blue dotted-dashed line) and cS-eSRRs (red dashed line) show the normal Lorentzian resonant profiles. The solid lines in Fig. 5 show the calculated phase difference on the observation points 1 (black) and 2 (green) on the cD-eSRR [see Fig. 4(a)]. At lower peak frequency (0.90 THz), both resonators TO and SO oscillate in phase, while the phase difference between the oscillations of the TO and the SO reaches a maximum value around 2.57 rad at the higher peak frequency (0.97 THz). These features agree well with the study of Fano resonance using two different meta-atoms.[12, 29] From these results, the meta-atoms with tight and sparse periodicities can be considered as the two-coupled oscillators.

In summary, we demonstrated coupled resonator metamaterials, where the meta-atoms are arranged with double periodicity. The double periodicity with sub-wavelength lattice period means that one sub-lattice cannot overlap the other sub-lattice by translational operation. In the measured and calculated transmission spectra, we observed Fano resonance induced by coupling between resonators with tight and sparse periodicities. Our proposed metamaterials can be considered as two-oscillators. The concept of meta-atoms with double periodicity can provide an additional degree of freedom for engineering coupled resonator metamaterials. The Fano resonance induced by the coupling between the different lattice periods can be used to implement multi-band resonances by slightly reducing the symmetry of meta-atom lattice structure. Since we can also realize the multi-band metamaterials by increasing the number of the harmonic oscillators, the combining with other designs of the Fano-like metamaterials (e.g. using the different meta-atoms, the PC-PIT, or the FGIT) has the great possibilities for the development of the multi-band metamaterials. The multi-band features are very useful in many application fields such as the perfect absorber using metamaterials.

**Acknowledgments**

Authors acknowledge Prof. Teshima, Prof. Wagata, Prof. Hayashi and Mr. Kawashima in Shinshu University for their supports in sample evaluation. This work was partially supported by JSPS KAKENHI Grant Number 22109003.

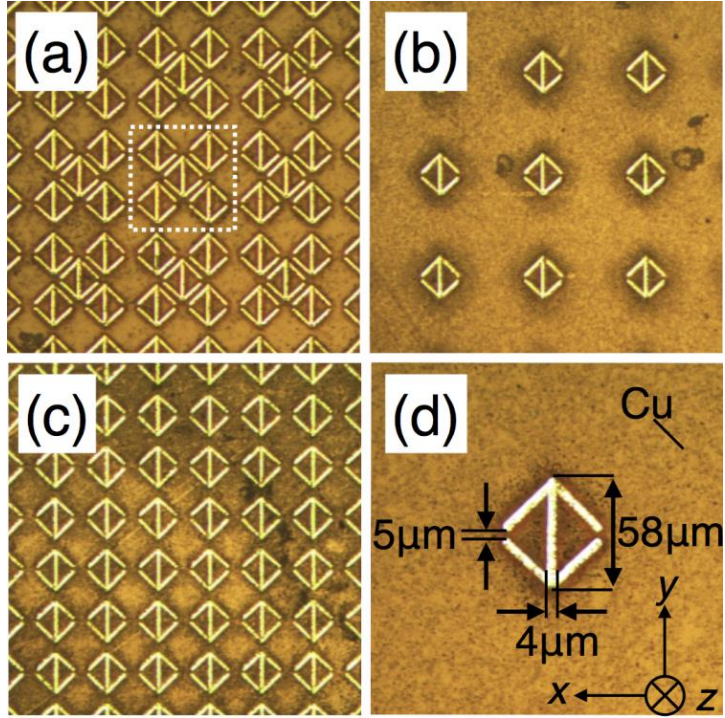

Fig. 1. Micrographs of (a) cD-eSRRs, (b) cS-eSRRs, and (c) cT-eSRRs. The region surrounded by white dashed line indicates the unit cell. (d) Geometrical parameters of a meta-atom. The incident electric field is directed to the $x$ direction, and the propagation direction is the $+z$ direction.

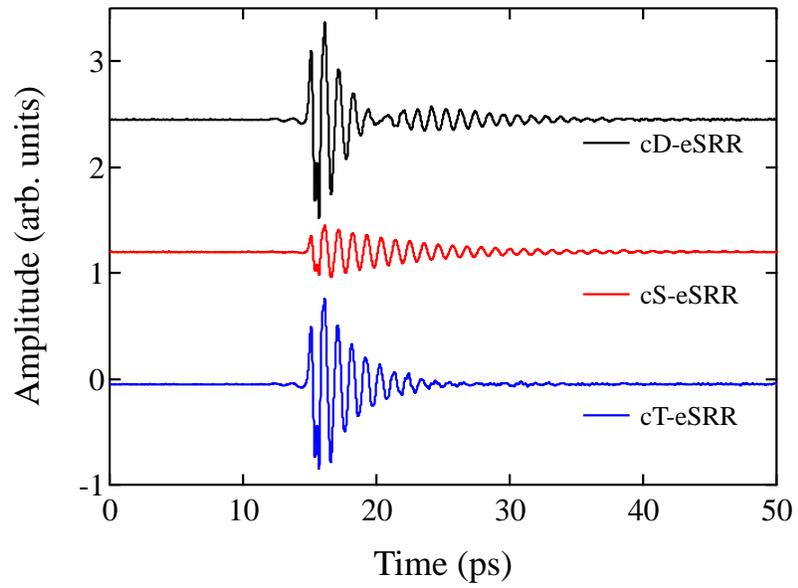

Fig. 2. Measured transmission waves of cT-eSRRs (blue), cS-eSRRs (red), and cD-eSRRs (black).

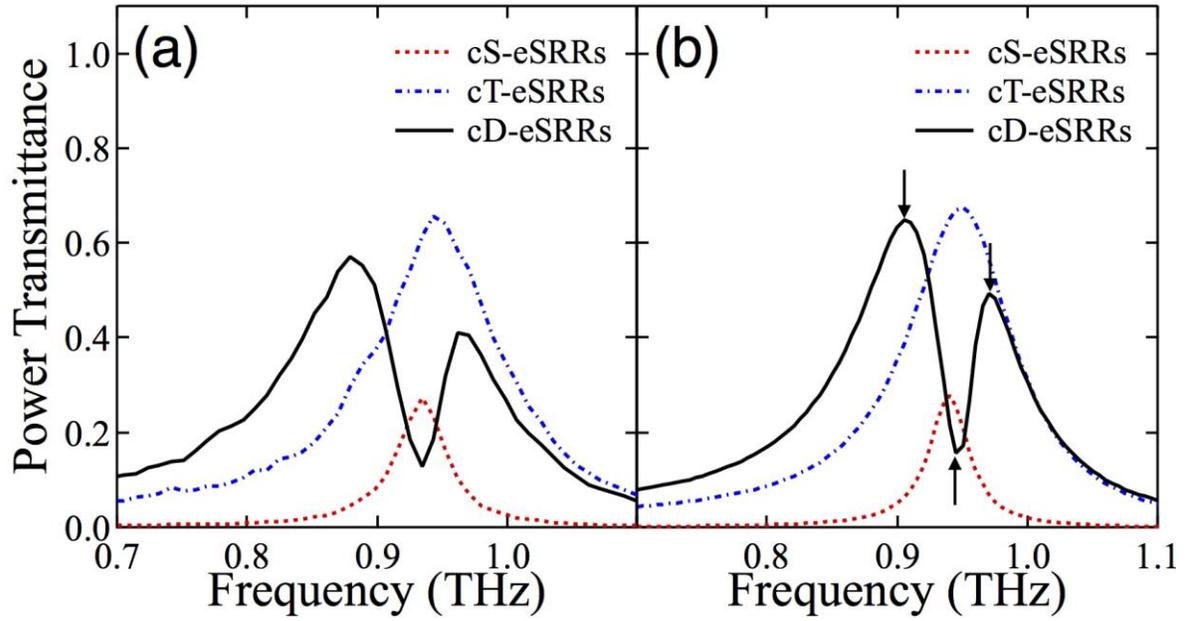

Fig. 3. (a) Measured and (b) simulated transmission spectra of cT-eSRRs (blue dotted-dashed lines), cS-eSRRs (red dashed lines), and cD-eSRRs (black solid line).

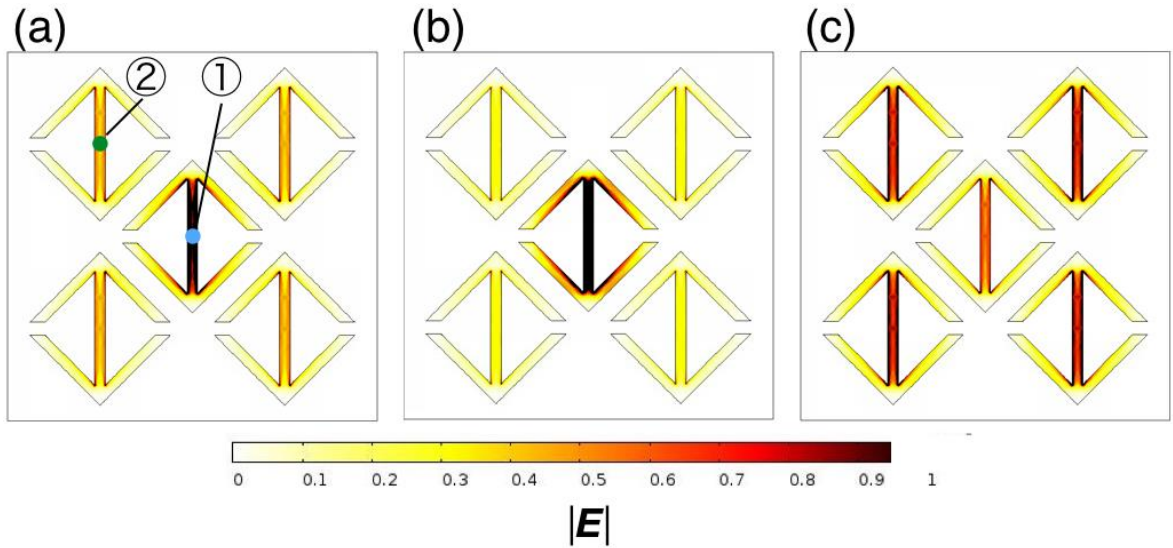

Fig. 4. Distributions of |$E$| of cD-eSRRs at (a) 0.90 THz, (b) 0.94 THz, and (c) 0.97 THz. The blue and green dots show the observation points to calculate the phase of the electric field.

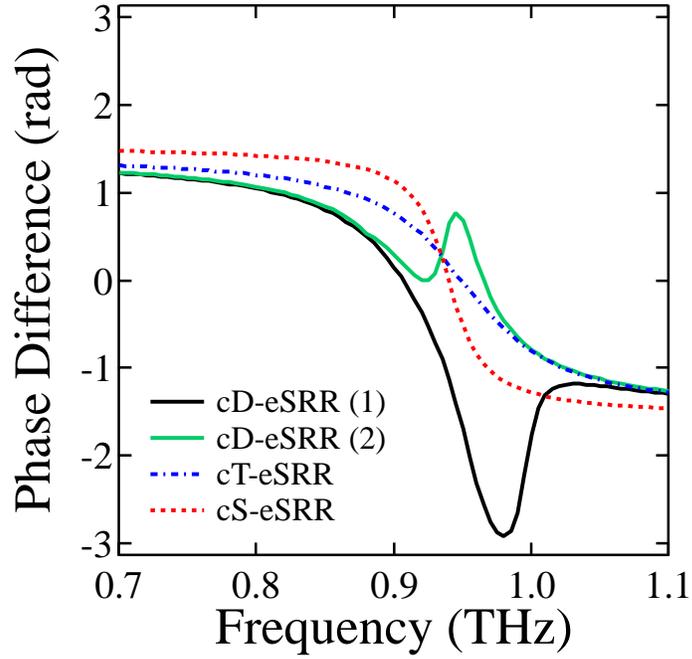

Fig. 5. The calculated phase differences of the electric field to the incident wave on the points 1 (black solid line) and 2 (green solid line) on the cD-eSRRs. The blue dotted-dashed line and the red dashed line show the phase difference on the center of cT-eSRRs and cS-eSRRs, respectively.